\documentclass[a4paper,12pt]{article}
\usepackage[utf8]{inputenc}
\usepackage[T1]{fontenc}
\usepackage{mathtools}
\usepackage{mathrsfs}
\usepackage{mathbbol}
\usepackage{amsthm}
\usepackage{amssymb}
\usepackage[top=0.91in, left=1in, right=1in, bottom=0.9in]{geometry}
\usepackage[margin=0.2\linewidth,labelfont=bf]{caption}
\usepackage{tikz}
\usepackage{transparent}
\usetikzlibrary{decorations.markings}
\usetikzlibrary{calc}
\usepackage{bigints}
\usepackage[
    sorting=none,
    style=numeric-comp,
    maxbibnames=99,
    date=year,
    doi=false,
    url=false,
    giveninits=true
  ]{biblatex}
\usepackage{aas_macros}
\usepackage[nodisplayskipstretch]{setspace}
\usepackage{microtype}
\usepackage{hyperref}
\usepackage[normalem]{ulem}
\usepackage{marginnote}
\usepackage{subcaption}
\usepackage{wrapfig}
\hypersetup{
	colorlinks,
	linkcolor={red!60!black},
	citecolor={blue!60!black},
	urlcolor={blue!80!black},
	breaklinks=true
}
\usepackage{physics}

\renewcommand*\footnoterule{{\color{white!60!black}\noindent}}
\let\oldfootnote\footnote
\renewcommand\footnote[1]{\oldfootnote{\setstretch{1.5}\hspace{.7mm}#1}}

\DeclareFieldFormat{doilink}{\iffieldundef{doi}{#1}{\href{http://dx.doi.org/\thefield{doi}}{#1}}}
\DeclareFieldFormat[article]{volume}{\textbf{#1}}
\DeclareFieldFormat[article]{pages}{#1}
\DeclareFieldFormat[inbook]{pages}{#1}
\DeclareFieldFormat[inproceedings]{pages}{#1}
\DeclareFieldFormat[inbook]{title}{\mkbibitalic{#1}}
\DeclareFieldFormat[inproceedings]{title}{\mkbibitalic{#1}}
\DeclareFieldFormat[article]{title}{\mkbibitalic{#1}}
\DeclareFieldFormat[unpublished]{title}{\mkbibitalic{#1}}
\DeclareBibliographyDriver{article}{%
  \usebibmacro{bibindex}%
  \usebibmacro{begentry}%
  \usebibmacro{author/translator+others}%
  \setunit{\labelnamepunct}\newblock
  \usebibmacro{title}%
  \newunit
  \printlist{language}%
  \newunit\newblock
  \usebibmacro{byauthor}%
  \newunit\newblock
  \usebibmacro{bytranslator+others}%
  \newunit\newblock
  \printfield{version}%
  \newunit\newblock
  \printtext[doilink]{%
  \usebibmacro{journal+issuetitle}%
  \newunit
  \usebibmacro{byeditor+others}%
  \newunit
  \usebibmacro{note+pages}%
  }%
  \newunit\newblock
  \iftoggle{bbx:isbn}
    {\printfield{issn}}
    {}%
  \newunit\newblock
  \usebibmacro{doi+eprint+url}%
  \newunit\newblock
  \usebibmacro{addendum+pubstate}%
  \setunit{\bibpagerefpunct}\newblock
  \usebibmacro{pageref}%
  \usebibmacro{finentry}
}

\def\diffD{\mathrm{D}}
\DeclareDocumentCommand\Differential{ o g d() }{ 
	\IfNoValueTF{#2}{
		\IfNoValueTF{#3}
		{\diffd\IfNoValueTF{#1}{}{^{#1}}}
		{\mathinner{\diffd\IfNoValueTF{#1}{}{^{#1}}\argopen(#3\argclose)}}
	}
	{\mathinner{\diffD\IfNoValueTF{#1}{}{^{#1}}#2} \IfNoValueTF{#3}{}{(#3)}}
}
\DeclareDocumentCommand\Dd{}{\Differential} 

\DeclareMathAlphabet\mathbfcal{OMS}{cmsy}{b}{n}
\DeclareMathAlphabet\mathbfscr{OMS}{mdugm}{b}{n}

\DeclareMathOperator*{\minext}{min\,ext}
\DeclareMathOperator*{\ext}{ext}

\newcommand{\OIST}{\raisebox{-0.08em}{\includegraphics[height=0.8em]{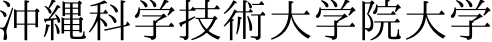}}}
\title{Probes, purviews, purgatories, parable, paradox?}
\author{Josh Kirklin}
\date{}

\bibliography{essayrefs}
\begin{document}
\begin{center}
  \linespread{1.55}\selectfont%
    \vspace*{1em}

    {\Large\bf Probes, purviews, purgatories, parable, paradox?}
    \vspace{1.75em}

    Josh Kirklin\\
    \url{joshua.kirklin@oist.jp}
    \vspace{1.75em}

    \emph{
        \OIST \\
        {\rm(}\!Okinawa Institute of Science and Technology{\rm)},\\
        1919-1 Tancha, Onna, Okinawa 904-0495, Japan
    }
    \vspace*{4em}

    \includegraphics[width=0.45\linewidth]{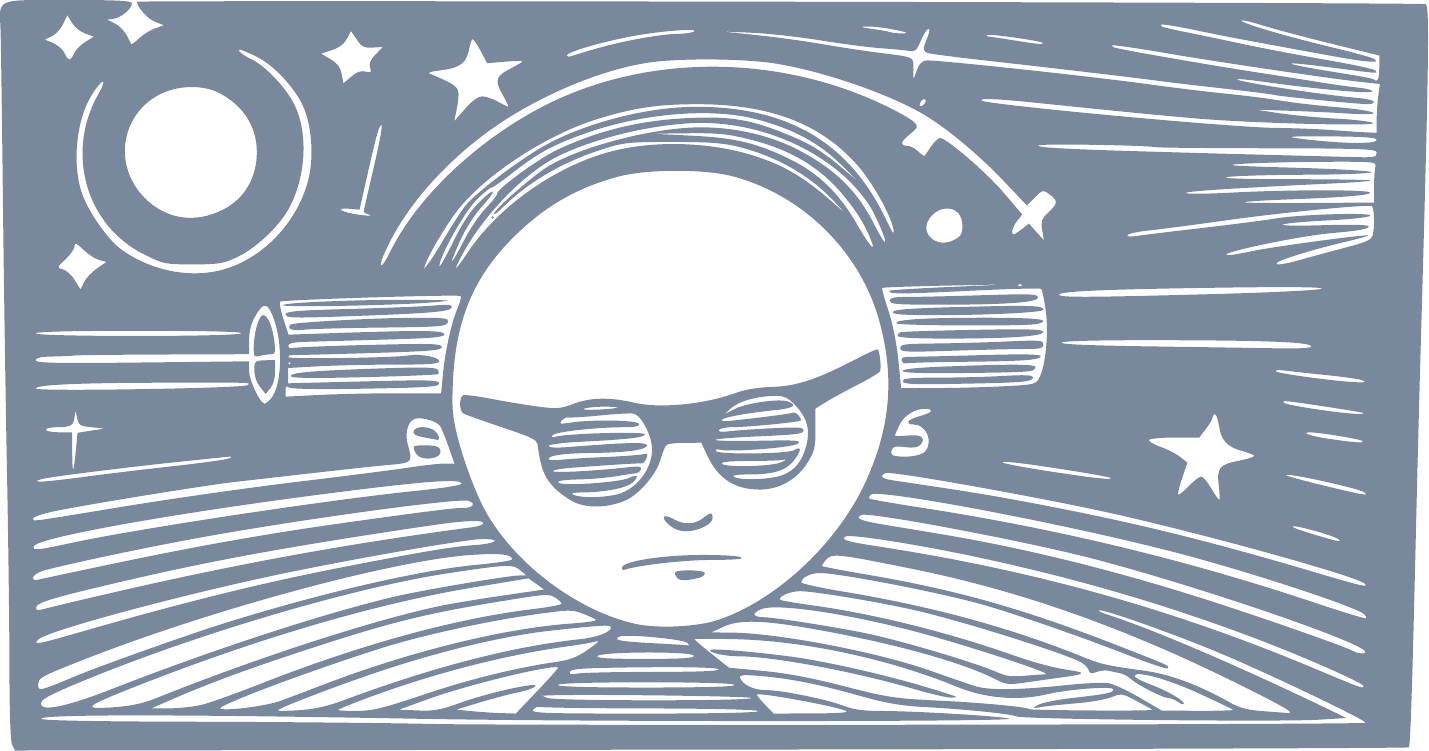}
    \vspace{1.5em}

    \begin{abstract}
        I discuss some general information-theoretic properties of quantum mechanical probes in semiclassical gravity: their purview, i.e.\ what they can see and act on (in terms of a generalised entanglement wedge), their spontaneous evaporation into a cloud of highly entropic particles when one tries to make them see too much (perhaps a parable on the dangers of straining one's eyes), and the subsequent resolution of an apparent information paradox.
    \end{abstract}
    \vfill

    {\sc\footnotesize Essay written for the Gravity Research Foundation \\2023 Awards for Essays on Gravitation}
    \vspace{1em}

    {\footnotesize Submitted: 30 March, 2023}
    \vspace{1.5em}
\end{center}
\newpage
\renewcommand*\footnoterule{{\color{white!60!black}\noindent\rule{\textwidth}{.4pt}}}

In this essay, I will discuss some information-theoretic properties of probes in semiclassical gravity. I will show: (\emph{i}) the entanglement entropy of a probe's internal degrees of freedom is given by the minimum extremum of the generalised entropies of all regions containing the probe; (\emph{ii}) the region with the minimal/extremal generalised entropy defines the probe's purview, i.e.\ what it can see, and what it can act on; (\emph{iii}) if we try to make the probe see too much, it will appear to spontaneously evaporate into a cloud of highly entropic particles; and (\emph{iv}) this mechanism prevents a paradox that could involve an accelerating probe violating an entropy bound.

Let me first explain how I will formulate such probes. The quantum gravity path integral $\mathcal{Z}$ involves a sum over spacetime topologies $\tau$, geometries $g$ and other degrees of freedom. In \emph{semiclassical} gravity, one assumes that the sum over $\tau,g$ is sharply peaked, so that when there is only a single dominant configuration (i.e.\ away from `phase transitions'), one can write
\begin{equation}
  \mathcal{Z} \approx \min_\tau \ext_g e^{iS_{\text{grav.}}[\tau,g]} \mathcal{Z}_{\phi}[\tau,g],
  \label{Equation: semiclassical path integral}
\end{equation}
where $\mathcal{Z}_{\phi}[\tau,g] = \int\Dd{\phi} e^{iS_{\text{fields}}[\tau,g,\phi]}$ is the path integral of an effective QFT on the curved background $\tau,g$, and the variable $\phi$ includes perturbative gravitons. The minimisation over $\tau$ is required for consistently viewing~\eqref{Equation: semiclassical path integral} as an approximation of an exact quantum theory. The rest of the formula accounts for all of semiclassical physics. 

To couple a quantum mechanical probe to quantum gravity, one could directly modify the path integral of the exact theory to include the probe, but the nature of such an approach would be highly constrained by the properties of the microscopic degrees of freedom. To stay UV-agnostic, I will instead modify~\eqref{Equation: semiclassical path integral}, changing $\mathcal{Z}_\phi$ to the path integral $\mathcal{Z}_{\phi,l,q}$ of an effective QFT coupled to an effective quantum mechanical probe, with $l$ being the embedding of the probe worldline in spacetime, and $q$ being its internal degrees of freedom. Each of $\phi,l,q$ are integrated over in $\mathcal{Z}_{\phi,l,q}$, but I will further assume that the integral over $l$ is sharply peaked, so that (away from phase transitions)
\begin{equation}
  \mathcal{Z} \approx \min_\tau \ext_{g,l} e^{iS_{\text{grav.}}[\tau,g] + iS_{\text{worldline}}[\tau,g,l]}  \mathcal{Z}_{\phi,q}[\tau,g,l],
  \label{Equation: semiclassical path integral with semiclassical probe}
\end{equation}
where $\mathcal{Z}_{\phi,q}$ is the effective path integral over only $\phi,q$. The properties of the probe can be tuned by appropriately choosing the worldline action $S_{\text{worldline}}$, and the form of $q$ in $\mathcal{Z}_{\phi,q}$.

\begin{figure}
  \centering
  \begin{subfigure}{0.3\linewidth}
    \centering
    \begin{tikzpicture}[thick, scale=0.7]
          \draw[blue!70, fill=blue!20] (0,0) .. controls (1,1.8) and (3,1.8) .. (4,0)
              coordinate[pos=0.65] (a)
            .. controls (3,-1.8) and (1,-1.8) .. (0,0)
              coordinate[pos=0.35] (b);

            \fill[red] (a) circle (0.1);
            \fill[red] (b) circle (0.1);
            \draw[red, decoration={
            markings,
            mark=at position 0.5 with {\arrow[scale=2]{<}}
          }, postaction={decorate}] (a) to[out=250, in=110, looseness=1.1] (b);

            \node[blue!70] at (1,1.6) {$\Sigma_+$};
            \node[blue!70] at (1,-1.6) {$\Sigma_-$};
            \node[red] at (3.2,1.45) {$x_+$};
            \node[red] at (3.2,-1.45) {$x_-$};

    \end{tikzpicture}
    \caption{}
    \label{Figure: possible topologies: basic}
  \end{subfigure}
  \begin{subfigure}{0.3\linewidth}
    \centering
    \begin{tikzpicture}[thick, scale=0.7]
          \draw[blue!70, fill=blue!20] (0,0) .. controls (1,1.8) and (3,1.8) .. (4,0)
              coordinate[pos=0.65] (a)
            .. controls (3,-1.8) and (1,-1.8) .. (0,0)
              coordinate[pos=0.35] (b);

            \fill[red] (a) circle (0.1);
            \fill[red] (b) circle (0.1);
            \draw[red, decoration={
            markings,
            mark=at position 0.5 with {\arrow[scale=2]{<}}
          }, postaction={decorate}] (a) to[out=250, in=110, looseness=1.1] (b);

            \node[blue!70] at (1,1.6) {$\Sigma_+$};
            \node[blue!70] at (1,-1.6) {$\Sigma_-$};
            \node[red] at (3.2,1.45) {$x_+$};
            \node[red] at (3.2,-1.45) {$x_-$};

            \draw[blue!40,fill=blue!15] (2,-3) circle (1);
            \shade[ball color = blue!40, opacity = 0.15] (2,-3) circle (1);
            \draw[blue!40,dashed] (1,-3) arc (180:0:1 and 0.3);
            \draw[blue!40] (1,-3) arc (180:360:1 and 0.3);
    \end{tikzpicture}
    \caption{}
    \label{Figure: possible topologies: spacetime closed component}
  \end{subfigure}
  \begin{subfigure}{0.3\linewidth}
    \centering
    \begin{tikzpicture}[thick, scale=0.7]
          \draw[blue!70, fill=blue!20] (0,0) .. controls (1,1.8) and (3,1.8) .. (4,0)
              coordinate[pos=0.65] (a)
            .. controls (3,-1.8) and (1,-1.8) .. (0,0)
              coordinate[pos=0.35] (b);

            \fill[red] (a) circle (0.1);
            \fill[red] (b) circle (0.1);
            \draw[red, decoration={
            markings,
            mark=at position 0.5 with {\arrow[scale=1.5]{<}}
          }, postaction={decorate}] (1.3,0) circle (0.4);
            \draw[red, decoration={
            markings,
            mark=at position 0.5 with {\arrow[scale=2]{<}}
          }, postaction={decorate}] (a) to[out=250, in=110, looseness=1.1] (b);

            \node[blue!70] at (1,1.6) {$\Sigma_+$};
            \node[blue!70] at (1,-1.6) {$\Sigma_-$};
            \node[red] at (3.2,1.45) {$x_+$};
            \node[red] at (3.2,-1.45) {$x_-$};

    \end{tikzpicture}
    \caption{}
    \label{Figure: possible topologies: worldline closed component}
  \end{subfigure}
  \caption{Some example spacetimes (in \textcolor{blue!70}{blue}) and probe worldlines (in \textcolor{red}{red}) that contribute to the path integral.}
  \label{Figure: possible topologies}
\end{figure}

The states which play a role in~\eqref{Equation: semiclassical path integral with semiclassical probe} are specified by a choice of conditions constraining (\emph{i})~the topology and geometry of the boundary of spacetime, (\emph{ii})~the configuration of the fields in the vicinity of the boundary, and (\emph{iii})~the location and internal state of the probe at any points it crosses the boundary. One evaluates the path integral $\mathcal{Z}$ over a range of configurations consistent with these conditions. Which kinds of topologies are allowed, and how regular the fields are required to be, is ultimately determined by the properties of the exact theory.\footnote{In the semiclassical approximation, configurations involving closed components of spacetime or the probe worldline, disconnected from the boundary on which the states are defined, like in Figures~\ref{Figure: possible topologies: spacetime closed component} and~\ref{Figure: possible topologies: worldline closed component}, can be dismissed on general grounds (assuming the probe does not backreact strongly on the spacetime) --- because such contributions will either be suppressed (relative to the same contribution with the closed components removed) by loop corrections to the path integral, or can be absorbed into a renormalisation of the effective theory.} Some examples are given in Figure~\ref{Figure: possible topologies}. 

Suppose $\mathcal{H}_{\text{grav.}}, \mathcal{H}_{\text{pr.}}$ and $\mathcal{H}_{\text{grav.}+\text{pr.}}$ are the effective Hilbert spaces of the quantum gravity system (including the fields $\phi$), the probe's internal degrees of freedom, and the combined gravity-probe system respectively. I shall assume the existence of an (approximate) isometry $V: \mathcal{H}_{\text{grav.$+$pr.}} \to \mathcal{H}_{\text{grav.}}\otimes \mathcal{H}_{\text{pr.}}$ which decomposes the combined gravity-probe state into those of the two individual systems.\footnote{\label{Footnote: V nonunique}The map $V$ depends on \emph{when} we choose to do this decomposition, and thus is not unique. One could define a whole family $V_\tau$ of such maps, each of which does the decomposition at different proper times $\tau$ along the probe's worldline; this would enable a description of the probe's dynamics. I will just assume we have made a fixed choice of $V$.} $\mathcal{H}_{\text{grav.$+$pr.}}$ is not the same as $\mathcal{H}_{\text{grav.}}\otimes\mathcal{H}_{\text{pr.}}$ because of gauge symmetry (the gauge constraints are modified in the presence of the probe); the assumption that $V$ nevertheless exists means that the probe must not strongly backreact on the gravitational background. Intuitively speaking, this map `excises' the probe from spacetime, and the lack of significant backreaction means this can be done without too much drama.\footnote{Note that there must be \emph{some} backreaction, to account for the changing constraints. This is related to the fact that $V$ is necessarily only an approximate isometry.}

$V$ may be used to study the entanglement between the probe and the gravitational system. For example, by acting with it on a joint state $\ket{\Psi}\in\mathcal{H}_{\text{grav.}+\text{pr.}}$, and then taking a partial trace over $\mathcal{H}_{\text{grav.}}$, one obtains a reduced state $\rho_{\text{pr.}} = \tr_{\text{grav.}}\qty(V\ket{\Psi}\bra{\Psi}V^\dagger)$ for the probe. I will now show how to compute the probe's entanglement entropy $\mathrm{S}_{\text{pr.}}=-\tr_{\text{pr.}}(\rho_{\text{pr.}}\log\rho_{\text{pr.}})$ using a replica trick and the semiclassical path integral~\eqref{Equation: semiclassical path integral with semiclassical probe}. The account is similar to e.g.~\cite{Ryu:2006bv,Lewkowycz:2013nqa,Engelhardt:2014gca}: one evaluates $\mathrm{S}_{\text{pr.}}$ as the $n\to 1$ limit of the R\'enyi entropy
\begin{equation}
  \mathrm{S}_n = \frac1{1-n}\tr_{\text{pr.}}\rho_{\text{pr.}}^n
  = \frac1{1-n}\tr_{\text{pr.}}\Big(\underbrace{\tr_{\text{grav.}}\qty(V\ket{\Psi}\bra{\Psi}V^\dagger)\dots \tr_{\text{grav.}}\qty(V\ket{\Psi}\bra{\Psi}V^\dagger)}_{n}\Big).
\end{equation}
In the path integral, for each $\ket{\Psi}$ that appears here, one sets up a boundary $\Sigma_-$ of spacetime containing an endpoint $x_-\in\Sigma_-$ of a probe worldline, and initial data for the fields and probe. For each $\bra{\Psi}$ one similarly sets up another boundary $\Sigma_+$ and probe endpoint $x_+\in\Sigma_+$, and initial data for them, in an orientation-reversed manner. This results in $n$ of each boundary and of each probe endpoint: $\Sigma_\pm^i$, $x_\pm^i$, $i=1,\dots,n$. 

\begin{figure}
  \centering
  \begin{tikzpicture}[thick, scale=0.7]
    \foreach \i in {0,1,2,3,4} {
      \begin{scope}[rotate={-\i*72}, shift={(-5,0)}]
        \draw[blue!70] (0,0) .. controls (1,0.8) and (3,0.8) .. (4,0)
            coordinate[pos=0.5] (a\i)
          .. controls (3,-0.8) and (1,-0.8) .. (0,0)
            coordinate[pos=0.5] (b\i);

          \pgfmathtruncatemacro{\si}{int(\i+1)}
          \node[blue!70] at (1,1) {$\Sigma^\si_+$};
          \node[blue!70] at (1,-1) {$\Sigma^\si_-$};
          \node[red] at (2.6,1) {\footnotesize$x^\si_+$};
          \node[red] at (2.6,-1) {\footnotesize$x^\si_-$};
      \end{scope}
    }
    \foreach \i in {0,1,2,3,4} {
      \pgfmathtruncatemacro{\j}{mod(\i+1,5)}
        \fill[red] (a\i) circle (0.1);
        \fill[red] (b\i) circle (0.1);
        \draw[red, decoration={
            markings,
            mark=at position 0.6 with {\arrow[scale=2]{<}}
          }, postaction={decorate}] (a\i) to[out={70-\i*72}, in={210-\i*72}, looseness=1.1] (b\j);
    }

    \draw[line width=2pt, {latex}-] (-30:6) arc (-30:30:6) node[midway, right] {\Large$\mathbb{Z}_n$};
  \end{tikzpicture}
  \caption{The boundaries and probe worldlines in the path integral for $\mathrm{\mathrm{S}}_n$ (here $n=5$). }
  \label{Figure: replica trick conditions}
\end{figure}

To account for the maps $V,V^\dagger$ and the partial traces, for each $i$ one should take the inner product between the gravity degrees of freedom on $\Sigma_+^i$ and $\Sigma_-^i$ (which means identifying their boundaries $\partial\Sigma_+^i$ and $\partial\Sigma_-^i$), and  (treating $i$ as an integer mod $n$) the probe degrees of freedom at $x_+^i$ and $x_-^{i+1}$ (which means drawing a worldline between those points\footnote{\label{Footnote: worldline conditions}Some additional conditions may need to be imposed on the worldline. For example, there may be constraints on its proper duration. This depends on which $V$ is being used (see footnote~\ref{Footnote: V nonunique}).}). This is illustrated in Figure~\ref{Figure: replica trick conditions}.
The spacetime boundary in the path integral for $\mathrm{S}_n$ consists of the glued together surfaces $\Sigma_i^\pm$, and the initial data for the fields and probe must be imposed appropriately on these surfaces. 

The boundaries and worldlines have a $\mathbb{Z}_n$ replica symmetry $i\to i+1$ (mod $n$) which I will assume is not spontaneously broken by the dominant minimal/extremal configuration of $\tau,g,l$. There are then basically two ways for spacetime to accommodate the worldlines: either the $n$ branches of spacetime are joined by wormholes to a single connected region used by all $n$ worldlines (Figure~\ref{Figure: possible replica topologies: one wormhole}), or each worldline traverses its own wormhole joining each pair of consecutive branches (Figure~\ref{Figure: possible replica topologies: n wormholes}).\footnote{The wormholes are prevented from pinching off to zero width by a version of the Casimir effect, which in~\eqref{Equation: entropy} manifests as a UV divergence in $\mathrm{S}_{\text{eff.}}$. If there were only an area term in~\eqref{Equation: entropy}, the minimal/extremal $\Sigma_{\text{pr.}}$ would have zero size. So entanglement in the effective theory is essential for the results given here.} I will call these the `alive' and `dead' phases respectively. There can also be additional wormholes cyclically connecting the branches, but not traversed by any worldlines (these lead to `islands' below); they have been omitted from the Figures. In a real time Lorentzian path integral, the metric and other fields take on complexified values inside the wormholes~\cite{Colin-Ellerin:2020mva}.

\begin{figure}
  \centering
  \begin{subfigure}{0.49\linewidth}
    \centering
    \begin{tikzpicture}[thick, scale=0.6]
      \foreach \i in {0,1,2,3,4} {
        \draw[blue!70,fill=blue!20,rotate={-\i*72}, shift={(-5.5,0)}] (0,0) .. controls (1,1.4) and (3,1.4) .. (4,0)
            coordinate[pos=0.35] (a\i)
          .. controls (3,-1.4) and (1,-1.4) .. (0,0)
            coordinate[pos=0.65] (b\i);
    }

    \foreach \i in {0,1,2,3,4} {
        \coordinate (p\i) at ({50+\i*72}:3);
        \coordinate (q\i) at ({72+\i*72}:0.7);
        \coordinate (r\i) at ({94+\i*72}:3);
      }
      \fill[blue!20] 
         (p0) .. controls (q0) .. (r0)
      -- (p1) .. controls (q1) .. (r1)
      -- (p2) .. controls (q2) .. (r2)
      -- (p3) .. controls (q3) .. (r3)
      -- (p4) .. controls (q4) .. (r4)
      -- (p0);
      \begin{scope}
    \clip 
         (p0) .. controls (q0) .. (r0)
      -- (p1) .. controls (q1) .. (r1)
      -- (p2) .. controls (q2) .. (r2)
      -- (p3) .. controls (q3) .. (r3)
      -- (p4) .. controls (q4) .. (r4)
      -- (p0);
    \foreach \i in {0,1,2,3,4} {
        \draw[blue!40,dotted,semithick,rotate={-\i*72}, shift={(-5.5,0)}] (0,0) .. controls (1,1.4) and (3,1.4) .. (4,0)
          .. controls (3,-1.4) and (1,-1.4) .. (0,0);
      }
      \end{scope}

      \coordinate (back) at (-0.05,0.15);
      \coordinate (front) at (-0.05,-0.15);
    \foreach \i in {2,3} {
      \pgfmathtruncatemacro{\j}{mod(\i+1,5)}
      \draw[green!70!black,densely dotted] ({72*\i}:1.2) to[out={72*\i-90}, in={72*\i}, looseness=0.5] (back);
    }
    \foreach \i in {0,1,4} {
      \pgfmathtruncatemacro{\j}{mod(\i+1,5)}
      \draw[green!70!black,densely dotted] ({72*\i}:1.2) to[out={72*\i+90}, in={72*\i}, looseness=0.5] (back);

    }
    \foreach \i in {0,1,2,3,4} {
      \draw[blue!40] (p\i) .. controls (q\i) .. (r\i);
    }
    \foreach \i in {2,3} {
      \pgfmathtruncatemacro{\j}{mod(\i+1,5)}
      \draw[green!60!black, line width = 1.75pt] ({72*\i}:1.2) to[out={72*\i+90}, in={72*\i}, looseness=0.5] (front);
    }
    \foreach \i in {0,1,4} {
      \pgfmathtruncatemacro{\j}{mod(\i+1,5)}
      \draw[green!60!black, line width = 1.75pt] ({72*\i}:1.2) to[out={72*\i-90}, in={72*\i}, looseness=0.5] (front);
    }
    \foreach \i in {2,3} {
      \pgfmathtruncatemacro{\j}{mod(\i+1,5)}
      \draw[green!80!black, line width = 0.75pt] ({72*\i}:1.2) to[out={72*\i+90}, in={72*\i}, looseness=0.5] (front);
    }
    \foreach \i in {0,1,4} {
      \pgfmathtruncatemacro{\j}{mod(\i+1,5)}
      \draw[green!80!black, line width = 0.75pt] ({72*\i}:1.2) to[out={72*\i-90}, in={72*\i}, looseness=0.5] (front);
    }
    \fill[green!60!black] (front) circle (0.1);

    \foreach \i in {0,1,2,3,4} {
      \pgfmathtruncatemacro{\j}{mod(\i+1,5)}
        \fill[red] (a\i) circle (0.1);
        \fill[red] (b\i) circle (0.1);
        \draw[red, decoration={
            markings,
            mark=at position 0.4 with {\arrow[scale=1.5]{<}}
          }, postaction={decorate}] (a\i) to[out={340-\i*72}, in={-50-\i*72}, looseness=3.4] (b\j);
        \begin{scope}[rotate={-\i*72}, shift={(-5,0)}]
          \pgfmathtruncatemacro{\si}{int(\i+1)}
          \node[blue!70] at (0,1.3) {$\Sigma^\si_+$};
          \node[blue!70] at (0,-1.3) {$\Sigma^\si_-$};
          \node[red] at (1.1,1.5) {\footnotesize$x^\si_+$};
          \node[red] at (1.1,-1.5) {\footnotesize$x^\si_-$};
        \end{scope}
    }

    \node[right,green!50!black] at (1.3,-0.1)  {\footnotesize$\Sigma_{\text{pr.}}$};
    \end{tikzpicture}
    \caption{}
    \label{Figure: possible replica topologies: one wormhole}
  \end{subfigure}
  \begin{subfigure}{0.49\linewidth}
    \centering
    \begin{tikzpicture}[thick, scale=0.6]
      \foreach \i in {0,1,2,3,4} {
      \begin{scope}[rotate={-\i*72}, shift={(-5.5,0)}]
        \draw[blue!70,fill=blue!20] (0,0) .. controls (1,1.4) and (3,1.4) .. (4,0)
            coordinate[pos=0.5] (a\i)
          .. controls (3,-1.4) and (1,-1.4) .. (0,0)
            coordinate[pos=0.5] (b\i);

          \coordinate (c\i) at (2.8,0.6);
          \coordinate (d\i) at (3.5,0.2);
          \coordinate (e\i) at (2.8,-0.6);
          \coordinate (f\i) at (3.5,-0.2);
      \end{scope}
    }
    \foreach \i in {0,1,2,3,4} {
      \pgfmathtruncatemacro{\j}{mod(\i+1,5)}
      \fill[blue!20] (c\i) -- (d\i) to[out={30-\i*72}, in={270-\i*72},looseness=0.8] (f\j) 
        -- (e\j) to[in={0-\i*72}, out={300-\i*72},looseness=0.8] (c\i);
    }
    \foreach \i in {0,1,2,3,4} {
      \pgfmathtruncatemacro{\j}{mod(\i+1,5)}
      \begin{scope}
          \clip (c\i) -- (d\i) to[out={30-\i*72}, in={270-\i*72},looseness=0.8] (f\j) 
            -- (e\j) to[in={0-\i*72}, out={300-\i*72},looseness=0.8] (c\i);
          \draw[blue!40,dotted,semithick,rotate={-\i*72}, shift={(-5.5,0)}] (0,0) .. controls (1,1.4) and (3,1.4) .. (4,0)
            .. controls (3,-1.4) and (1,-1.4) .. (0,0);
          \draw[blue!40,dotted,semithick,rotate={-72-\i*72}, shift={(-5.5,0)}] (0,0) .. controls (1,1.4) and (3,1.4) .. (4,0)
            .. controls (3,-1.4) and (1,-1.4) .. (0,0);
      \end{scope}
  }
    \foreach \i in {0,1,2,3,4} {
      \pgfmathtruncatemacro{\j}{mod(\i+1,5)}
      \draw[blue!40] (c\i) to[out={0-\i*72}, in={300-\i*72},looseness=0.8] (e\j);
      \draw[blue!40] (d\i) to[out={30-\i*72}, in={270-\i*72},looseness=0.8] (f\j);
        \begin{scope}[rotate={36-\j*72}, shift={(-5.5,0)}]
          \draw[green!70!black, densely dotted] (3.99,0) arc (360:180:0.29 and 0.1);
          \draw[green!60!black, line width = 1.75pt] (3.99,0) arc (0:180:0.29 and 0.1);
          \draw[green!80!black, line width = 0.75pt] (3.99,0) arc (0:180:0.29 and 0.1);
        \end{scope}
      }
    \foreach \i in {0,1,2,3,4} {
      \pgfmathtruncatemacro{\j}{mod(\i+1,5)}
        \fill[red] (a\i) circle (0.1);
        \fill[red] (b\i) circle (0.1);
        \draw[red, decoration={
            markings,
            mark=at position 0.45 with {\arrow[scale=1.5]{<}}
          }, postaction={decorate}] (a\i) to[out={300-\i*72}, in={0-\i*72}, looseness=2.7] (b\j);
        \begin{scope}[rotate={-\i*72}, shift={(-5,0)}]
          \pgfmathtruncatemacro{\si}{int(\i+1)}
          \node[blue!70] at (0,1.3) {$\Sigma^\si_+$};
          \node[blue!70] at (0,-1.3) {$\Sigma^\si_-$};
          \node[red] at (1.6,1.6) {\footnotesize$x^\si_+$};
          \node[red] at (1.6,-1.6) {\footnotesize$x^\si_-$};
        \end{scope}
      }

      \node[left,green!50!black] at (1.7,-0.1)  {\footnotesize$\Sigma_{\text{pr.}}$};
    \end{tikzpicture}
    \caption{}
    \label{Figure: possible replica topologies: n wormholes}
  \end{subfigure}
  \caption{The replica spacetime in \textbf{(a)}~the `alive' phase, \textbf{(b)}~the `dead' phase. The surfaces which become $\Sigma_{\text{pr.}}$ after quotienting by $\mathbb{Z}_n$ are highlighted in \textcolor{green!50!black}{green}.}
  \label{Figure: possible replica topologies}
\end{figure}

By a well-known construction~\cite{Ryu:2006bv,Lewkowycz:2013nqa,Engelhardt:2014gca} one may take a `quotient' over the $\mathbb{Z}_n$ replica symmetry to write this $n$-fold path integral in terms of only a single copy of the system (with only a single spacetime boundary $\Sigma_-\cup\Sigma_+$ and a single worldline), which enables one to more easily take the limit $n\to 1$; the resulting expression for the probe entropy is
\begin{equation}
  \mathrm{S}_{\text{pr.}}=\lim_{n\to 1}\mathrm{S}_n = \minext_{\Sigma_{\text{pr.}}}\qty( \frac{\operatorname{Area}[\partial\Sigma_{\text{pr.}}]}{4G} + \mathrm{S}_{\text{eff.}}[\Sigma_{\text{pr.}}]),
  \label{Equation: entropy}
\end{equation}
where the ``$\minext$'' ranges over all codimension 1 regions $\Sigma_{\text{pr.}}$ in spacetime which intersect the probe worldline (these regions come from cutting up the wormholes in the $n$-fold replica spacetime), and $\mathrm{S}_{\text{eff.}}[\Sigma_{\text{pr.}}]$ is the entanglement entropy of the effective degrees of freedom in $\Sigma_{\text{pr.}}$ (the full term in parentheses is known as the `generalised entropy' of $\Sigma_{\text{pr.}}$). The right-hand side should be evaluated on the dominant configuration of $\tau,g,l$ for the single set of boundary conditions and worldline.

\begin{wrapfigure}{R}{0.29\linewidth}
    \centering
    \begin{tikzpicture}[thick,scale=1]
          \draw[blue!70, fill=blue!20] (0,0) .. controls (1,1.8) and (3,1.8) .. (4,0)
              coordinate[pos=0.65] (a)
            .. controls (3,-1.8) and (1,-1.8) .. (0,0)
              coordinate[pos=0.35] (b);

            \fill[red] (a) circle (0.1);
            \fill[red] (b) circle (0.1);

            \draw[green!50!black,fill=green!30,shift={(0.1,0)}] (2.2,0.5) -- (2.7,0) -- (2.2,-0.5) -- (1.7,0) -- cycle;
            \draw[green!50!black,fill=green!30,scale=0.7,shift={(-1,0)}] (2.2,0.5) -- (2.7,0) -- (2.2,-0.5) -- (1.7,0) -- cycle;

            \draw[red, decoration={
                markings,
                mark=at position 0.5 with {\arrow[scale=2]{<}}
              }, postaction={decorate}] (a) to[out=250, in=110, looseness=1.1] (b);

            \node[blue!70] at (1,1.55) {$\Sigma_+$};
            \node[blue!70] at (1,-1.55) {$\Sigma_-$};
            \node[red] at (3.2,1.45) {$x_+$};
            \node[red] at (3.2,-1.45) {$x_-$};
        \end{tikzpicture}
        \caption{An example entanglement wedge (in \textcolor{green!50!black}{green}) for the probe, including an island.}
    \label{Figure: example entanglement wedge}
\end{wrapfigure}
Thus, the entanglement entropy of the probe is given by the minimum/extremum of the generalised entropy of $\Sigma_{\text{pr.}}$, taken over all possible regions $\Sigma_{\text{pr.}}$ intersecting its worldline.\footnote{Note that the worldline and boundary conditions are invariant under complex conjugation and time reversal. Assuming this symmetry is not spontaneously broken in the path integral, it may be shown that the minimal/extremal $\Sigma_{\text{pr.}}$ will be near to the probe proper time specified by $V$ (see footnotes~\ref{Footnote: V nonunique} and~\ref{Footnote: worldline conditions}).} $\Sigma_{\text{pr.}}$ can contain components disconnected from the probe, a.k.a.\ islands. By analogy with holography, I will refer to the domain of dependence of the minimal/extremal $\Sigma_{\text{pr.}}$ as the `entanglement wedge' of the probe. An example is given in Figure~\ref{Figure: example entanglement wedge}.
Actually, a series of arguments similar to those in~\cite{Jafferis:2015del,Dong:2016eik} shows furthermore that the \emph{relative} entropy of two nearby probe states is equal to the relative entropy of the effective degrees of freedom in the entanglement wedge. This then guarantees (through the theorems of operator algebra quantum error correction~\cite{Harlow:2016vwg,Faulkner:2020hzi}) the existence of an injective unital $*$-homomorphism $R:\mathcal{A}_{\text{E.W.}}\to \mathcal{B}(\mathcal{H}_{\text{pr.}})$ such that $R(\mathcal{O})V = V \mathcal{O}$ for all $\mathcal{O}\in\mathcal{A}_{\text{E.W.}}$, where $\mathcal{A}_{\text{E.W.}}$ is the algebra of effective observables in the entanglement wedge, and $\mathcal{B}(\mathcal{H}_{\text{pr.}})$ is the algebra of bounded operators acting on $\mathcal{H}_{\text{pr.}}$ (technically these statements only hold approximately). 

With $R$, one can measure and act on the entanglement wedge, using only the internal degrees of freedom of the probe. Also, $\mathcal{A}_{\text{E.W.}}$ is the largest algebra such that this is possible.
Let me step back slightly, to take stock of what this intuitively means. In general, a probe is used by first coupling it to a physical system, and then measuring the state of the probe. This procedure allows one to indirectly observe the system, but only some of its properties --- namely, those to which the probe has access. With this in mind, the interpretation of the above result should be clear: the probes constructed in this essay allow one to measure their entanglement wedges, and no more. In other words, the entanglement wedge is `what the probe can see' (and act on), a.k.a.\ its purview.

A probe's entanglement wedge always contains the probe's own neighbourhood, so it can see itself and the nearby effective fields. It can also see the fields in the islands, if there are any; in this sense, it has a `crystal ball' which can be used to look into far away, possibly causally disconnected regions, and act on the effective degrees of freedom contained in them. These islands will be present if and only if the effective degrees of freedom near the probe are sufficiently highly entangled with those far away, such that $\mathrm{S}_{\text{eff.}}$ competes with the area term in the generalised entropy. This could happen, for example, if the probe happens to collect a Page time's worth of Hawking radiation from a black hole; then it would have access to an island inside the black hole (see~\cite{Almheiri:2019hni,Penington:2019npb}).

The probe could also just be deliberately set up in a highly entangled state with some other effective degrees of freedom, so that it has an island containing them (this is a manifestation of $\text{ER} = \text{EPR}$~\cite{Maldacena:2013xja}). 
If Alice and Bob are two probes whose internal degrees of freedom are sufficiently highly entangled with one another, then Alice will have an island containing Bob's internal degrees of freedom, and Bob will have an island containing Alice's internal degrees of freedom, and they can use this to communicate with each other (this is a version of a quantum teleportation protocol~\cite{Bennett:1992tv}). Moreover, if their mutual information is sufficiently high, then their joint entanglement wedge will be larger than their individual entanglement wedges. In other words, they see more by working together. It would be interesting to investigate what happens when one of the probes is inside a black hole, and the other outside.

\makeatletter      
\newcommand{\getAngle}[2]{%
    \pgfmathanglebetweenpoints{\pgfpointanchor{#1}{center}}
                              {\pgfpointanchor{#2}{center}}
    \global\let\myangle\pgfmathresult 
}
\makeatother
\begin{figure}
  \centering
  \begin{tikzpicture}[scale=0.8]
      \draw[line width=1pt,-{latex}] (7.3,-3) node[below] {\small initial time} -- (7.3,1) node[midway,right] {$t$} node[above] {\small later time};

    \begin{scope}[thick, scale=0.6]
        \node at (3.8,-7.2) {\small\textbf{(a)}};

        \node[blue!70] at (-.5,-5) {$\Sigma_-$};
        \node[blue!70] at (7.7,-3.9) {$\Sigma_+$};
        \node[red] at (1.2,-5.5) {\footnotesize$x_-$};
        \node[red] at (6.1,-5) {\footnotesize$x_+$};

        \coordinate (left) at (0,0);
        \coordinate (right) at (7,0.2);
        \coordinate (front) at (1,-5);
        \coordinate (back) at (6,-4.5);
        \coordinate (top) at (3.5,2);

        \coordinate (leftmid) at ($(left)!0.2!(right)$);
        \coordinate (rightmid) at ($(left)!0.8!(right)$);
        \getAngle{left}{right}

        \fill[blue!20] (left) -- (leftmid) to[out=\myangle, in={180+\myangle}] (back)
            to[out={\myangle}, in={180+\myangle}] (right)
            to[out={180+\myangle},in={0}] (top)
            to[out={180},in=\myangle] (left);
        \draw[blue!70] (left) -- (leftmid) to[out=\myangle, in={180+\myangle}] (back)
            to[out={\myangle}, in={180+\myangle}] (right);
        \draw[red] (back) to[out=110,in=0,looseness=0.6] (top);

        \fill[blue!20] (left) to[out=\myangle, in={180-\myangle/3}] (front)
            to[out={-\myangle/3}, in={180+\myangle}] (rightmid) -- (right)
            to[out={180+\myangle},in={0}] (top)
            to[out={180},in=\myangle] (left);

        \begin{scope}
            \clip (left) to[out=\myangle, in={180-\myangle/3}] (front)
                to[out={-\myangle/3}, in={180+\myangle}] (rightmid) -- (right)
                to[out={180+\myangle},in={0}] (top)
                to[out={180},in=\myangle] (left);
            \draw[blue!70,dotted] (left) -- (leftmid) to[out=\myangle, in={180+\myangle}] (back)
                to[out={\myangle}, in={180+\myangle}] (right);
            \draw[red,dashed] (back) to[out=110,in=0,looseness=0.6] (top);
        \end{scope}

        \draw[red, decoration={
            markings,
            mark=at position 0.45 with {\arrow[scale=1.5]{>}}
          }, postaction={decorate}] (front) to[out=70,in=180,looseness=0.5] (top);
        \draw[blue!40] (left) to[out=\myangle, in={180}] (top)
            to[out={0}, in={180+\myangle}] (right);
        \draw[blue!70] (left) to[out=\myangle, in={180-\myangle/3}] (front)
            to[out={-\myangle/3}, in={180+\myangle}] (rightmid) -- (right);

        \fill[red] (back) circle (0.1);
        \fill[red] (front) circle (0.1);
    \end{scope}

    \begin{scope}[thick, scale=0.6, shift={(17,0)}]
        \node at (3.8,-7.2) {\small\textbf{(b)}};

        \node[blue!70] at (-.5,-5) {$\Sigma_-$};
        \node[blue!70] at (7.7,-3.9) {$\Sigma_+$};
        \node[red] at (1.2,-5.5) {\footnotesize$x_-$};
        \node[red] at (6.1,-5) {\footnotesize$x_+$};

        \coordinate (left) at (0,0);
        \coordinate (right) at (7,0.4);
        \coordinate (front) at (1,-5);
        \coordinate (back) at (6,-4.5);
        \coordinate (top) at (3.5,2);
        \coordinate (node) at (3.5,-1.7);

        \coordinate (leftmid) at ($(left)!0.35!(right)$);
        \coordinate (rightmid) at ($(left)!0.7!(right)$);
        \getAngle{left}{right}

        \fill[blue!20] (left) -- (leftmid) to[out=\myangle, in={180+\myangle}] (back)
            to[out={\myangle}, in={180+\myangle}] (right)
            to[out={180+\myangle},in={0}] (top)
            to[out={180},in=\myangle] (left);
        \draw[blue!70] (left) -- (leftmid) to[out=\myangle, in={180+\myangle}] (back)
            to[out={\myangle}, in={180+\myangle}] (right);
       \draw[red] (back) to[out=100, in=290] (4.9,-2.3) to[out=110,in=0] (node);
        \draw[red, decoration={
            markings,
            mark=at position 0.45 with {\arrow[scale=1.5]{>}}
          }, postaction={decorate}] (front) to[out=80,in=180] (node);
        \fill[blue!20] (2.2,-2.6) .. controls (2.3,-1.7) and (5.0,-1.7) .. (5.2,-2.6)
            coordinate[pos=0.2] (openbottom) coordinate[pos=0.7] (openbottomback)
            -- (4.7,-0.7) .. controls (4.6,-1.3) and (2.5,-1.3) .. (2.4,-0.7)
            coordinate[pos=0.85] (opentop) coordinate[pos=0.2] (opentopback);

        \fill[blue!20] (left) to[out=\myangle, in={180-\myangle/3}] (front)
            to[out={-\myangle/3}, in={180+\myangle}] (rightmid) -- (right)
            to[out={180+\myangle},in={0}] (top)
            to[out={180},in=\myangle] (left);
        \draw[blue!40,dotted] (opentopback) to[out=180,in=180,looseness=0.5] (openbottomback);
        \begin{scope}
            \clip (2.2,-2.6) .. controls (2.3,-1.7) and (5.0,-1.7) .. (5.2,-2.6) --
            (4.7,-0.7) .. controls (4.6,-1.3) and (2.5,-1.3) .. (2.4,-0.7);
            \draw[blue!70,dotted] (left) -- (leftmid) to[out=\myangle, in={180+\myangle}] (back)
                to[out={\myangle}, in={180+\myangle}] (right);
        \end{scope}
        \begin{scope}
            \clip (opentop) to[out=0,in=0,looseness=0.5] (openbottom) -- (5,-5.5) -- (5,0);
            \clip (2.2,-2.6) .. controls (2.3,-1.7) and (5.0,-1.7) .. (5.2,-2.6)
                -- (4.7,-0.7) .. controls (4.6,-1.3) and (2.5,-1.3) .. (2.4,-0.7);
            \draw[red, dashed] (back) to[out=100, in=290] (4.9,-2.3) to[out=110,in=0] (node) to[out=180,in=80] (front);
        \end{scope}
        \draw[blue!40] (2.2,-2.6) .. controls (2.3,-1.7) and (5.0,-1.7) .. (5.2,-2.6);
        \draw[blue!40] (2.4,-0.7) .. controls (2.5,-1.3) and (4.6,-1.3) .. (4.7,-0.7);
        \begin{scope}
            \clip (left) to[out=\myangle, in={180-\myangle/3}] (front)
                to[out={-\myangle/3}, in={180+\myangle}] (rightmid) -- (right)
                to[out={180+\myangle},in={0}] (top)
                to[out={180},in=\myangle] (left);
            \draw[blue!70,dotted] (left) -- (leftmid) to[out=\myangle, in={180+\myangle}] (back)
                to[out={\myangle}, in={180+\myangle}] (right);
            \begin{scope}
                \clip (opentop) to[out=0,in=0,looseness=0.5] (openbottom) -- (2,-5.5) -- (0,-5.5) -- (0,0);
                \draw[red, decoration={
                    markings,
                    mark=at position 0.45 with {\arrow[scale=1.5]{>}}
                  }, postaction={decorate}] (front) to[out=80,in=180] (node);
            \end{scope}
            \begin{scope}
                \clip (opentop) -- (openbottom) -- (6,-5) -- (6,1);
                \draw[blue!40,dotted, thin] (2.2,-2.6) .. controls (2.3,-1.7) and (5.0,-1.7) .. (5.2,-2.6);
                \draw[blue!40,dotted, thin] (2.4,-0.7) .. controls (2.5,-1.3) and (4.6,-1.3) .. (4.7,-0.7);
            \end{scope}
        \end{scope}

        \draw[blue!40] (openbottom) to[out=180,in=180,looseness=0.5] (opentop);

        \begin{scope}
            \clip (opentop) to[out=0,in=0,looseness=0.5] (openbottom) -- (2,-5.5) -- (0,-5.5) -- (0,0);
            \draw[red, decoration={
                markings,
                mark=at position 0.45 with {\arrow[scale=1.5]{>}}
              }, postaction={decorate}] (front) to[out=80,in=180] (node);
        \end{scope}

        \draw[blue!40] (left) to[out=\myangle, in={180}] (top)
            to[out={0}, in={180+\myangle}] (right);
        \draw[blue!70] (left) to[out=\myangle, in={180-\myangle/3}] (front)
            to[out={-\myangle/3}, in={180+\myangle}] (rightmid) -- (right);
    \centering

        \draw[blue!40] (opentop) to[out=0,in=0,looseness=0.5] (openbottom);
        \begin{scope}
            \clip (opentop) -- (openbottom) -- (2,-5) -- (0,-5) -- (0,1) -- (3,1);
            \draw[blue!40] (2.2,-2.6) .. controls (2.3,-1.7) and (4.8,-1.7) .. (4.9,-2.6);
            \draw[blue!40] (2.4,-0.7) .. controls (2.5,-1.3) and (4.6,-1.3) .. (4.7,-0.7);
        \end{scope}
        \draw[blue!40] (opentopback) to[out=0,in=0,looseness=0.5] (openbottomback);

        \fill[red] (back) circle (0.1);
        \fill[red] (front) circle (0.1);
    \end{scope}
  \end{tikzpicture}
  \caption{The fate of the probe in \textbf{(a)}~the alive phase, \textbf{(b)}~the dead phase.}
  \label{Figure: quotient spacetime}
\end{figure}

One can grow (or shrink) a probe's purview by modifying the state $\ket{\Psi}\in\mathcal{H}_{\text{grav.}+\text{pr.}}$. The most obvious way to do this would be to increase the entanglement between the effective degrees of freedom inside the current purview, and those just outside it. If there is a sufficiently high amount of new entanglement, then the purview would be enlarged --- as the decrease in $\mathrm{S}_{\text{eff.}}(\Sigma_{\text{pr.}})$ would overpower the increase in $\operatorname{Area}(\partial\Sigma_{\text{pr.}})/4G$.

But a warning is in order: this may also change whether the probe is in the alive phase or the dead phase. In Figure~\ref{Figure: quotient spacetime}, the spacetimes for the two phases have been drawn as `timefolds', to emphasise the flow of Lorentzian time $t$ (c.f.\ the Schwinger-Keldysh formalism). In the alive phase, the probe can observe the original universe in which it started. But in the dead phase, it splits off into a baby universe, and is useless for observing the original universe (assuming its purview contains no islands). A highly mixed effective QFT state is pasted in to the spacetime subregion the probe used to occupy --- so, from the perspective of the original universe, the probe appears to have spontaneously evaporated into a cloud of highly entropic particles. From its own point of view, the probe is not really dead; it lingers on in its purgatorial baby universe. In this case, $\Sigma_{\text{pr.}}$ loops around the entire baby universe,\footnote{Note that this means $\partial\Sigma_{\text{pr.}}$ is empty, so $\mathrm{S}_{\text{pr.}}$ will not contain an area term contribution.} so the `dead' probe's purview is its purgatory.

Suppose $\operatorname{dim}(\mathcal{H}_{\text{pr.}})$ is finite, so that the maximum possible probe entropy is $\mathrm{S}_{\text{pr.}} = \log(\dim(\mathcal{H}_{\text{pr}}))$. Since $\mathrm{S}_{\text{eff.}}\ge 0$, there is a bound on the size of probe's purview in the alive phase, in terms of its surface area:\footnote{Clearly the probe needs to have very many internal degrees of freedom, if one wants it to have access to a region much larger than the Planck scale. But note that so far I have only discussed the use of the probe at some fixed time on its worldline (in this sense the purview described here is only an \emph{instantaneous} one). If one were permitted to smear probe observables over some interval of the probe's time evolution, then one could access a larger spacetime region --- the causal completion of the union of the instantaneous purviews of all points in the corresponding interval of the worldline. If islands are present in any of the instantaneous purviews, they will be present in this smeared purview too (so this region can be larger than that obtained in~\cite{Witten:2023qsv}).}
\begin{equation}
    \operatorname{Area}(\partial\Sigma_{\text{pr.}}) \le 4G \log(\dim(\mathcal{H}_{\text{pr.}})).
    \label{Equation: maximum purview}
\end{equation}
What happens if a probe in the alive phase has a purview near its maximum size~\eqref{Equation: maximum purview}, but one still attempts to make it larger by building up entanglement with effective degrees of freedom in the exterior? The volume of the purview would have to increase, but its surface area $\operatorname{Area}(\partial\Sigma_{\text{pr.}})$ clearly cannot, so some curvature would need to be produced near the probe (assuming that the geometry of the purview is reasonably flat to start with).
There are two ways this can happen: (\emph{i}) A little bit of energy-momentum is placed near the probe, sourcing the curvature. But, assuming the probe and effective fields are only very weakly coupled together, there is nothing preventing this energy-momentum from radiating away from the vicinity of the probe, so the purview will quickly shrink back to its original size. (\emph{ii}) The topology of the region near the probe is changed. The only thing it can really change to is the dead phase. 

In some ways, the second option is an extreme version of the first: so much energy-momentum is introduced that a black hole forms (indeed, the dead phase may be understood as the probe falling into a newly-formed black hole which subsequently evaporates into Hawking radiation). It may also be viewed as a consequence of the monogamy of entanglement: the extra degrees of freedom added to the purview must be disentangled from, and hence topologically separated from, the rest of spacetime.

Thus, if one tries to increase the probe's purview past its maximum size, either the purview quickly shrinks back down again, or the probe is accidentally killed. So, a parable: \emph{don't strain your eyes, lest they spontaneously evaporate}.

The death of probes is not a complete tragedy, as it prevents a potential paradox. Suppose the probe is accelerating in an approximately flat spacetime, such that it is exposed to thermal Unruh radiation. This will cause the entanglement entropy of the effective degrees of freedom in its purview to steadily rise. If one insists on the probe staying in the alive phase, then there is no reason for this process to stop, and the probe's entropy would eventually violate the bound $\mathrm{S}_{\text{pr.}}\le \log(\dim(\mathcal{H}_{\text{pr.}}))$ --- a clear contradiction. The problem is resolved if the probe is allowed to die: it will then stop being exposed to Unruh radiation (since the Unruh effect is a flat spacetime phenomenon, and the probe is now confined to a non-Minkowski baby universe), so its entropy will stop growing.

The situation is analogous to the black hole information paradox, but with a probe instead of a black hole, and Unruh radiation instead of Hawking radiation. In both cases, there are information inconsistencies if one only accounts for one phase of spacetime (for the probe the `alive' phase, for black holes the so-called `Hawking' phase); in both cases, the violation is resolved by accounting for a novel phase of spacetime (here the `dead' phase, for black holes the `island' phase). It is fascinating that the topological properties of the semiclassical gravity path integral have such important consequences, even in the absence of exotic objects like black holes.

\setstretch{1.1}
\section*{Acknowledgements}
Thank you to Stefan Eccles, Christophe Goeller, Philipp H\"ohn, Juan Maldacena, Fabio Mele, and Malcolm Perry for helpful discussions and comments. This work was supported by funding from the Okinawa Institute of Science and Technology. 
This work was made possible through the support of the ID\# 62312 grant from the John Templeton Foundation, as part of the project \href{https://www.templeton.org/grant/the-quantum-information-structure-of-spacetime-qiss-second-phase}{``The Quantum Information Structure of Spacetime'' (QISS)}. The opinions expressed in this work are those of the author and do not necessarily reflect the views of the John Templeton Foundation.
The front image was created with the help of Stable Diffusion.

\setstretch{1.0}
\printbibliography

\end{document}